\documentclass[
pre,
showpacs,
preprint,
amsmath,
amssymb
]
{revtex4}
\usepackage[dvips]{graphicx}
\usepackage[dvips]{color}
\usepackage{amsmath}
\usepackage{amssymb}
\usepackage{dcolumn}
\usepackage{bm}
\begin{document}
\title{
Reentrant transition in the shear viscosity of dilute rigid rod dispersions
}

\author{Hideki Kobayashi}
\email{hidekb@cheme.kyoto-u.ac.jp}
\author{Ryoichi Yamamoto}
\email{ryoichi@cheme.kyoto-u.ac.jp}
\affiliation{Department of Chemical Engineering, Kyoto University, Kyoto
615-8510, Japan}
\affiliation{CREST, Japan Science and Technology Agency, Kawaguchi 332-0012, Japan}

\date{\today}

\begin{abstract}

The intrinsic viscosity of a dilute dispersion of rigid rods is studied
using a recently developed direct numerical simulation (DNS) method for
particle dispersions. 
A reentrant transition from shear-thinning to the 2nd Newtonian regime
is successfully reproduced in the present DNS results around a Peclet
number ${\rm Pe}=150$, which is in good agreement with our theoretical
prediction of ${\rm Pe}=143$, at which the dynamical crossover from
Brownian to non-Brownian behavior takes place in the rotational motion
of the rotating rod. 
The viscosity undershoot is observed in our simulations before reaching
the 2nd Newtonian regime. 
The physical mechanisms behind these behaviors are analyzed in detail.

\end{abstract}

\pacs{83.50.Ax, 83.60.Fg, 83.80.Rs, 47.57.Ng}

\maketitle

\section{INTRODUCTION}

The viscous properties of dilute dispersions of rigid rods change
drastically as the rate of applied shear flows $\dot{\gamma}$
increases. 
Although many previous studies have investigated this phenomenon, the
mechanism of this viscosity change is not yet completely clear. 
The aim of this paper is to understand the detailed mechanism of the
viscosity change by performing direct numerical simulations (DNS) for a
dilute dispersion of rigid rods that are subject to thermal fluctuations
in a Newtonian host fluid.

The relationship between the measurable bulk rheological properties and
the microscale description of dispersions of rod-like particles has been
previously investigated in the literature \cite{jeffrey, peterlin, giesekus, batchelor_I, Hinch_Leal_factor, Leal_Hinch_K, Hinch_Leal_theory1, Hinch_Leal_theory2, Leal_Hinch_review}. 
Giesekus obtained the expression for the bulk stress tensor of diluted
spheroidal dispersions under shear flow by taking into account the
effects of the rotational Brownian motion of the spheroids due to
thermal fluctuations \cite{giesekus}. 
Leal and Hinch reported that the viscosity behavior is characterized by
the aspect ratio $l$ of the rod and dimensionless shear rate
$\dot{\gamma}/D_{\rm r}$, where $D_{\rm r}$ is the rotational diffusion
constant \cite{Hinch_Leal_factor, Leal_Hinch_K, Hinch_Leal_theory1,
Hinch_Leal_theory2}.

In the case of weak shear flow, $\dot{\gamma}/D_{\rm r} \ll 1$, the
dilute rigid rod dispersions exhibit the 1st Newtonian behavior, in
which the viscosity $\eta$ of the dispersion is constant and equal to
the 1st Newtonian (zero-shear limiting) value $\eta_0$. 
For an intermediate regime, $1\ll\dot{\gamma}/D_{\rm r} \ll l^3 +
l^{-3}$, the dispersions exhibit shear-thinning behavior, in which
$\eta\propto(\dot{\gamma}/D_{\rm r})^{-1/3}$. 
In the case of strong shear flow, $l^3 + l^{-3} \ll \dot{\gamma}/D_{\rm
r}$, the dispersions reenter the 2nd Newtonian regime, in which $\eta$
becomes constant again and is equal to the 2nd Newtonian (high-shear
limiting) value $\eta_\infty$. 
Similar results have also been obtained in numerical \cite{sheraga,
shear_thinning_sim1} and experimental \cite{rod_experiment1,
rod_experiment2} studies. 
In this paper, the phrase {\it viscosity transition} is used to express
the changes in viscosity from the 1st Newtonian to the shear-thinning
behavior and also from the shear-thinning to the 2nd Newtonian
behavior. 
Similar results have been observed for dilute dispersions of flexible
chains, both experimentally \cite{exp1_transition} and theoretically
\cite{exp_theo_transition}.

Hinch and Leal \cite{Hinch_Leal_factor, Leal_Hinch_K} proposed a theoretical model for
the viscosity transitions. 
They considered that the viscosity $\eta$ of the dispersion is
determined by the ensemble average of the temporal viscosity
$\hat{\eta}(\theta,\varphi)$ using the probability distribution function
(PDF) $P_{\dot{\gamma}}(\theta,\varphi)$ of the two orientational angles
$\theta$ and $\varphi$ of the rod, {\it i.e.},
\begin{equation}
\eta(\dot{\gamma})=\int \hat{\eta}(\theta,\varphi)
P_{\dot{\gamma}}(\theta,\varphi) d\theta d\varphi.
\label{eta}
\end{equation}
Here the form of $P_{\dot{\gamma}}(\theta,\varphi)$ is shear rate
dependent, and the shear rate dependence of the dispersion viscosity
$\eta(\dot{\gamma})$ is introduced mainly through this function.

The rigid rod undergoes a random rotational Brownian motion at low shear
rates in the 1st Newtonian regime, where the effect of thermal
fluctuations is dominant over the effect of shear flow. 
Therefore,
\begin{equation}
P_{\dot{\gamma}}(\theta,\varphi)=constant
\end{equation}
holds over the entire phase space of $\theta$ and $\varphi$. 
The
viscosity is thus constant with respect to the shear rate change in this
regime, {\it i.e.}, $\eta(\dot{\gamma})=\eta_0$.

In contrast, the rigid rod undergoes a deterministic tumbling motion due
to strong shear flow in the 2nd Newtonian regime. 
Here the tumbling motion is perfectly described by Jeffery's equation
\cite{jeffrey}. 
Therefore, the PDF approaches the high-shear limiting (non-Brownian)
asymptotic form with increasing $\dot{\gamma}$,
\begin{equation}
P_{\dot{\gamma}}(\theta,\varphi)=P_J (\theta, \varphi),
\end{equation}
where $P_J(\theta, \varphi)$ is the
theoretical result \cite{Leal_Hinch_K} derived from Jeffery's equation \cite{jeffrey}. 
The viscosity, therefore,
tends to be constant again in this regime, {\it i.e.}, $\eta=\eta_\infty$.

\begin{figure}[htb]
\begin{center}
\includegraphics[width=0.7\hsize]{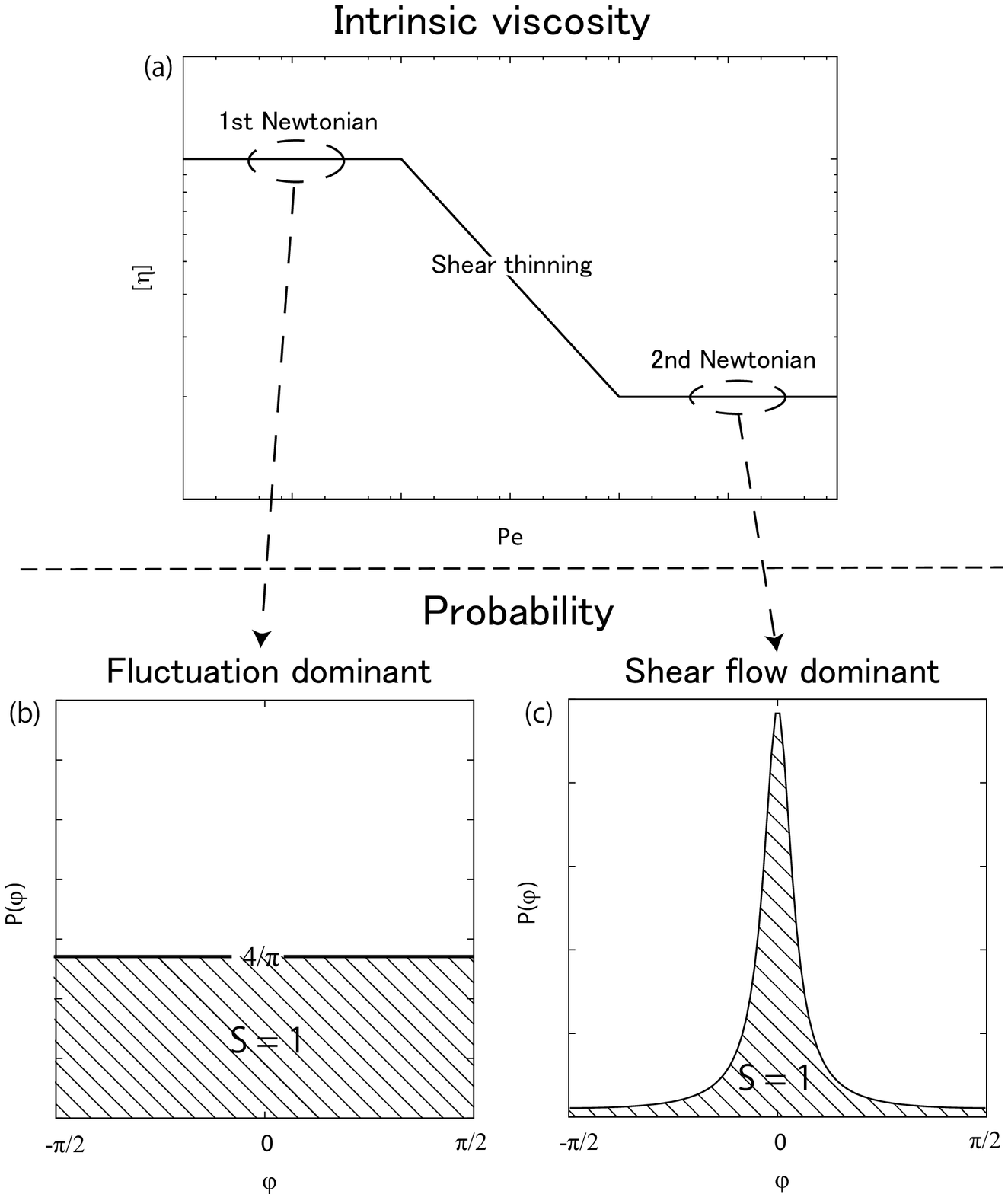}
\caption{A schematic illustration of the viscosity transition. (a) A typical behavior of intrinsic viscosity $[\eta]$ as a function of {Pe}. Here, $P(\varphi)\equiv\int_{-\frac{\pi}{2}}^{\frac{\pi}{2}} \cos\theta P(\theta,\varphi) d\theta$ is normalized so that $S\equiv\int_{-\frac{\pi}{2}}^{\frac{\pi}{2}}P(\varphi) d\varphi=1$. (b) $P(\varphi)$ in the weak-shear regime where the rod undergoes random tumbling. (c) $P(\varphi)$ in the strong shear regime where the rod undergoes periodic tumbling.}
\label{intro}
\end{center}
\end{figure}

The viscosity exhibits strong shear-thinning behavior in the
intermediate regime. 
The PDF is approximately given by
\begin{equation}
P_{\dot{\gamma}}(\theta,\varphi) \simeq P_J(\theta, \varphi) + (D_{\rm r}/\dot{\gamma}) P_{1}(\theta,\varphi), 
\end{equation}
where $P_{1}$ represents the leading term of the perturbation expansion
of the thermal effects. 
It is clearly seen that the contribution from the thermal effects
decreases with as the dimensionless shear rate $\dot{\gamma}/D_r$
increases in this regime, which gives rise to drastic shear-thinning
behavior. 
The solid line that is shown in three different flow regimes in
Fig.~\ref{intro} represents a schematic illustration of the viscosity
transition based on the above considerations.

Consistent with the theoretical model of Hinch and Leal
\cite{Hinch_Leal_factor, Leal_Hinch_K}, the viscosity transition from the 1st Newtonian
to shear-thinning regimes has already been successfully reproduced in
various numerical studies \cite{shear_thinning_sim1,
shear_thinning_sim2}. 
However, the viscosity transition from the shear-thinning to the 2nd
Newtonian regime has never been successfully reproduced by numerical
simulations. 
For rigid rod dispersions, we could not find any previous studies that
have been performed at high enough shear rates to approach the 2nd
Newtonian regime. 
Several numerical simulations have been conducted for flexible chain
dispersions at high shear rates that are expected to be in the 2nd
Newtonian regime. 
However, the viscosity transition from shear-thinning to the 2nd
Newtonian behavior has never been correctly reproduced, not even when
the hydrodynamic interactions are taken into account using the
Rotne-Prager-Yamakawa (RPY) tensor \cite{shear_thinning_sim2}.

In the present study, we used a different class of approach, called the
smoothed profile method (SPM) \cite{spm, spm2, spm_fl, spm_fl2,
oblique}, that can accurately take into account the thermal fluctuations
and the hydrodynamic coupling between bead particles with a finite
radius $a$ and a Newtonian host fluid, based on direct numerical
simulations (DNS) of particle dispersions. 
The viscosity of a rigid rod dispersion has been calculated using SPM to
reproduce the viscosity transition from shear-thinning to the 2nd
Newtonian regimes and to examine carefully the validity of the
theoretical model proposed by Hinch and Leal \cite{Hinch_Leal_factor, Leal_Hinch_K}.

\section{METHODS}

\subsection{Model}

We solve the dynamics of a single rigid rod in a Newtonian solvent using
SPM \cite{spm, spm2, spm_fl, spm_fl2}. 
In this method, the boundaries between solid particles and solvents are
replaced with a continuous interface by assuming a smoothed profile. 
This simple modification enables us to calculate the hydrodynamic
interactions both efficiently and accurately, without neglecting
many-body interactions. 
The equation governing a solvent with density $\rho_{\rm f}$ and shear
viscosity $\eta_{\rm f}$ is a modified Navier-Stokes equation:
\begin{equation}
  \rho_{\rm f}  \left[ \frac{\partial {\bf u}}{\partial t}+ 
		 ({\bf u}\cdot{\bm \nabla}){\bf u} \right] =
  -{\bm \nabla}p + \eta_{\rm f} {\bm \nabla}^2{\bf u} +
  \rho_{\rm f}\phi{\bf f}_{\rm p} + {\bf f}_{\rm shear}
 \end{equation}
with the incompressible condition ${\bm \nabla}\cdot{\bf u}=0$, where
${\bf u}({\bf r},t)$ and $p({\bf r},t)$ are the velocity and pressure
fields of the solvent, respectively. 
A smoothed profile function $0\leq\phi({\bf r},t)\leq1$ distinguishes
between the fluid and particle domains, yielding $\phi=1$ in the
particle domain and $\phi=0$ in the fluid domain. 
These domains are separated by thin interstitial regions, the
thicknesses of which are given by $\xi$. 
The body force $\phi{\bf f}_{\rm p}$ is introduced to ensure the
rigidity of the particles and the appropriate non-slip boundary
condition at the fluid/particle interface. 
The mathematical expressions for $\phi$ and $\phi{\bf f}_{\rm p}$ are
detailed in our previous papers \cite{spm,spm2}. 
The external force ${\bf f}_{\rm shear}$ is introduced to maintain a
linear shear with a shear rate of $\dot{\gamma}$. 
This force is applied with the oblique coordinate transformation based
on tensor analysis \cite{onuki,oblique}.

In the present study, we use a bead-spring model consisting of $N$
freely rotating beads in a single rigid rod. 
The bead diameter is $\sigma=2a$. 
The motion of the $i$th bead is governed by the following Newton-Euler
equations of motion with thermal fluctuations:
\begin{equation}
  M_i\frac{d}{dt}{\bf V}_i={\bf F}_i^{\rm H}+{\bf F}_i^{\rm
   P}+{\bf F}_i^C+{\bf G}_i^V,\;\;\;
   \frac{d}{dt}{\bf R}_i={\bf V}_i,
\end{equation}
\begin{equation}
  {\bf I}_i\cdot\frac{d}{dt}{\bm \Omega}_i={\bf N}_i^{\rm H}+{\bf G}_i^{\Omega},
\end{equation}
where ${\bf R}_i$, ${\bf V}_i$, and ${\bm \Omega}_i$ are the position,
translational velocity, and rotational velocity of the beads,
respectively. 
$M_i$ and ${\bf I}_i$ are the mass and moment of inertia, and ${\bf
F}_i^{\rm H}$ and ${\bf N}_i^{\rm H}$ are the hydrodynamic force and
torque exerted by the solvent on the beads, respectively
\cite{spm,spm2}. 
${\bf G}_i^{\rm V}$ and ${\bf G}_i^{\Omega}$ are the random force and
torque, respectively, due to thermal fluctuations. 
The temperature of the system is defined such that the long-time
diffusive motion of dispersed particles reproduces the correct behavior
\cite{spm_fl,spm_fl2}. 
${\bf F}_i^{\rm P}$ represents the potential force due to direct
inter-bead interactions.

We use a bead-spring model as a model of rod-like objects with a
truncated Lennard-Jones potential and a finitely extensible nonlinear
elastic (FENE) potential. 
The truncated Lennard-Jones interaction is expressed in terms of $U_{\rm LJ}$:
\begin{eqnarray}
   U_{\rm LJ}(r_{ij})=\left\{
		   \begin{array}{lll}
		    4\epsilon \left[
			       \left( \dfrac{\sigma}{r_{ij}}\right)^{12} -
			       \left( \dfrac{\sigma}{r_{ij}}\right)^{6}\right]
		    + \epsilon &(r_{ij}<2^{\frac{1}{6}}\sigma) \\
		    0 &(r_{ij}>2^{\frac{1}{6}}\sigma),
		   \end{array}
			\right.
\end{eqnarray}
where $r_{ij}=|{\bf R}_i - {\bf R}_j|$. 
The parameter $\epsilon$ characterizes the strength of the interactions,
and $\sigma$ represents the diameter of the beads. 
Consecutive beads on a chain are connected by a FENE potential of the form
\begin{equation}
U_{\rm FENE}(r)=-\frac{1}{2}k_{\rm c}R_0^2\ln\left[1-\left(\frac{r}{R_0}\right)^2\right],
\end{equation}
where $r=|{\bf R}_{i+1} - {\bf R}_i|$, $k_c=30\epsilon/\sigma^2$, and
$R_0=1.5\sigma$.
${\bf F}_i^{\rm C}$ is the constraint force acting on the $i$th bead due
to the bond-angle constraints that cause the connected beads to form a
straight rod.
\begin{equation}
  {\bf F}_i^{\rm C}=\frac{\partial}{\partial{\bf R}_i}(\displaystyle
   \sum^N_{\alpha=3} {\bm \mu}_{\alpha}\cdot{\bm \Psi}_{\alpha}),
\end{equation}
\begin{equation}
  {\bm \Psi}_{\alpha}=(\alpha-2){\bf R}_1-(\alpha-1){\bf R}_2+{\bf R}_\alpha,
   \label{constraint}
\end{equation}
where ${\bm \Psi}_{\alpha}=0$ is the constraint condition to be
satisfied. 
${\bm \mu}_{\alpha}$ is a Lagrange multiplier associated with the
constraints that is chosen such that the condition ${\bm
\Psi}_{\alpha}=0$ is satisfied at a time $t+h$, where $h$ is the time
increment of a single simulation step.

\begin{figure}[htb]
\begin{center}
\includegraphics[width=0.7\hsize]{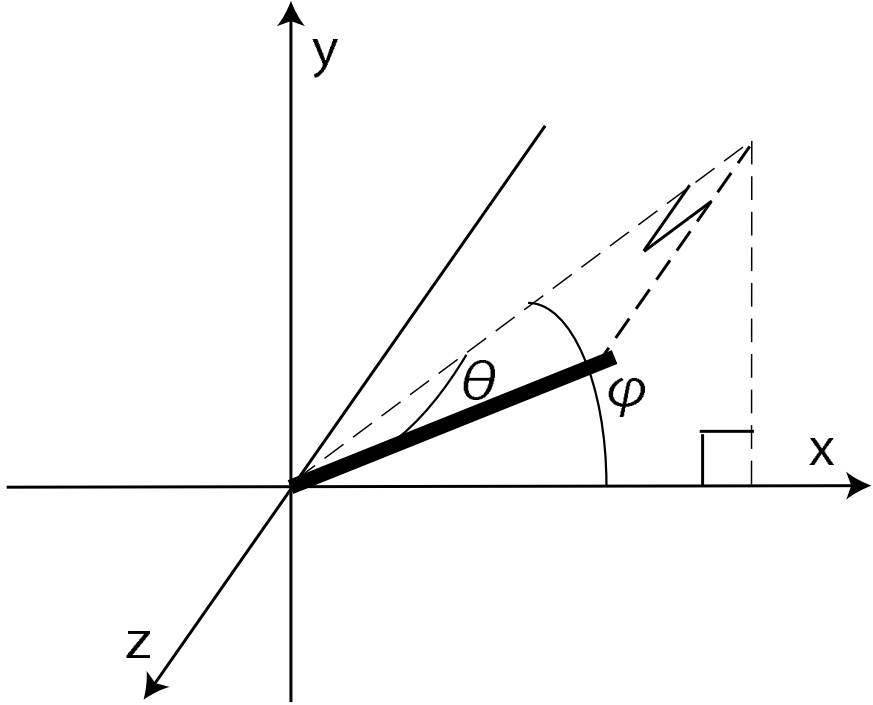}
\caption{The geometry of the rod's orientation in the present simulations.}
\label{geometry}
\end{center}
\end{figure}

The numerical simulations are performed in three dimensions with
periodic boundary conditions. 
The lattice spacing $\Delta$ is taken to be the unit of length. 
The unit of time is given by $\rho_{\rm f}\Delta^2/\eta_{\rm f}$, where
$\eta_{\rm f}=1$ and $\rho_{\rm f}=1$. 
The system size is $L_x\times L_y\times L_z=32\times32\times32$. 
The other parameters include the following: $\sigma=4$, $\xi=2$,
$\epsilon=1$, $M_i=4\pi a^3/3$, $N=5$, and $h=6.7\times10^{-2}$. 
In the presented simulations under shear flow, the Navier-Stokes
equation is discretized with a de-aliased Fourier spectral scheme in
space and with an Euler scheme in time \cite{oblique}. 
To follow the motions of the beads, the positions, velocities and
angular velocities of the beads are integrated with the Adams-Bashforth
scheme. 
The bead particles are assumed to be neutrally buoyant, so no gravity
effects are considered. 
At $t=0$, the rigid rod aligns along the $x$-axis, which is the flow
direction. 
The total duration $\tau_{\rm t}$ of each simulation is set such that
$\dot{\gamma}\tau_{\rm t}\simeq 3500$. 
The range of $k_{\rm B}T$ is $5.0\times10^{-4}<k_{\rm B}T<32$ and that
of $\dot{\gamma}$ is $5.0\times10^{-3}<\dot{\gamma}<2.0\times10^{-2}$. 
From the symmetry of the system, we follow the polar angles $\theta$ and
$\varphi$ defined in Fig.~\ref{geometry} to consider the motion of a
rigid rod. 
The angle defined between the rod and the $x$-$y$ plane is denoted by
$\theta$, and the angle defined between the rod projected on the $x$-$y$
plane and the $x$-axis is denoted by $\varphi$.

\subsection{Effective Aspect Ratio}

\begin{figure}[htb]
\begin{center}
\includegraphics[width=1.0\hsize]{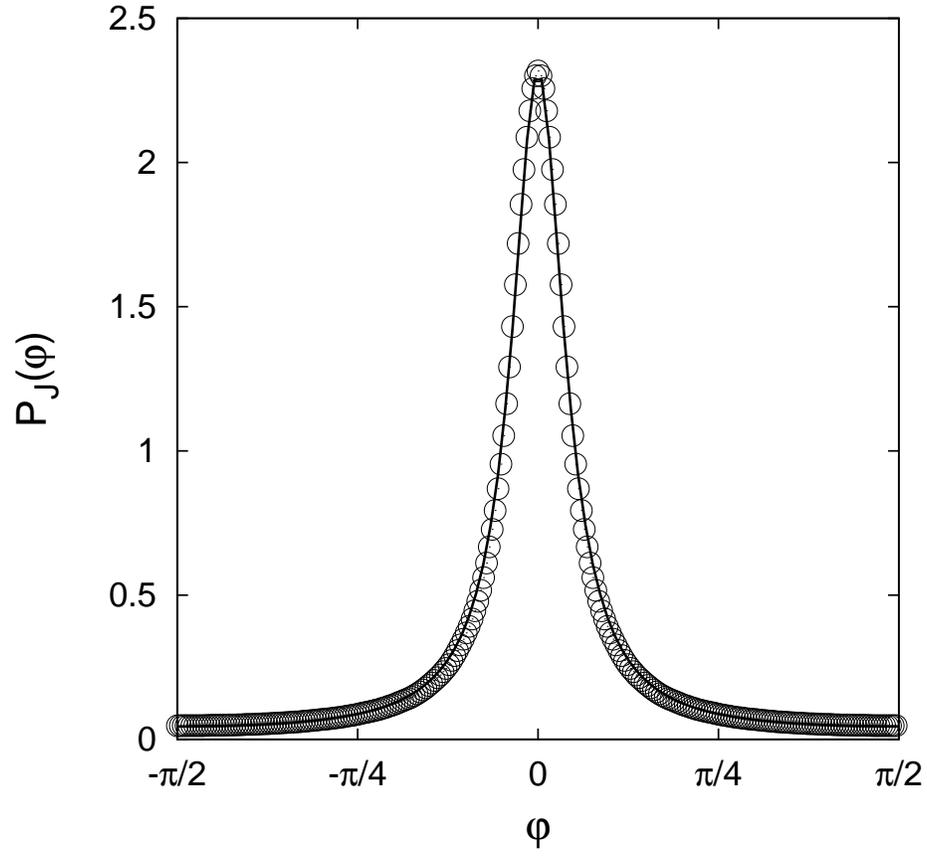}
\caption{The PDF for a rotational rigid rod as a function of $\varphi$ without thermal fluctuations. Numerical results (open circle) and Eq.~(\ref{p_j}) with $l=7.1$ (solid line).}
\label{probability}
\end{center}
\end{figure}

In the present study, the rigid rod is represented as connected beads. 
Because the beads composing the rod can rotate freely, the effective
aspect ratio $l$ differs from the simple geometrical aspect ratio
$L/\sigma$, where $L\simeq N\sigma$ is the rod's length. 
Instead, we evaluate $l$ numerically with the PDF of the rotating rigid
rods without thermal fluctuations in the $x$-$y$ plane, {\it i.e.},
$\theta=0$, as represented by
\begin{equation}
  P_{\rm J}(\varphi)= \frac{C_{0}}{\frac{l^2-1}{l^2+1}\sin^2 \varphi +
   \frac{1}{l^2+1}},
   \label{p_j}
\end{equation}
where $C_0$ is determined from the normalization condition
 $\int_{-\frac{\pi}{2}}^{\frac{\pi}{2}} P_{\rm J}(\varphi) d\varphi=1$
 \cite{jeffrey}. 
One obtained Eq.~(\ref{p_j}) in the following manner. 
The projection of the PDF $P_{\rm J}(\varphi)$ of a rotating rigid rod
on the $x$-$y$ plane is governed by a Fokker-Planck equation of the
form
\begin{equation}
\frac{\partial P_{\rm J}(\varphi)}{\partial t} =  \frac{\partial (\omega P_{\rm J}(\varphi))}{\partial \varphi} +
  2D_{\rm r}\frac{ \partial P_{\rm J}(\varphi)}{\partial \varphi^{2}},
  \label{fokker-planck}
\end{equation}
where $\omega=\dot{\varphi}$ is the angular velocity of the tumbling
rod. 
When the rigid rod rotates in the $x$-$y$ plane without thermal
fluctuations in steady states, the Fokker-Planck equation is modified to
\begin{equation}
  \frac{\partial (\omega P_{\rm J}(\varphi))}{\partial \varphi} = 0.
   \label{non_thermal_steady_fp}
\end{equation}
In this case, $\omega$ is represented as
\begin{equation}
  \omega  = \dot{\gamma} \left(
			  \frac{l^2-1}{l^2+1}\sin^2 \varphi + \frac{1}{l^2+1}
			 \right)
  \label{omega_velocity}
\end{equation}
from Jeffery's equation \cite{jeffrey}. 
Eq.~(\ref{p_j}) is obtained because $P_{\rm J}(\varphi)$ is inversely
proportional to $\omega$. 
Figure~\ref{probability} shows that our numerical results of
$P'(\varphi)=\int \cos \theta P_{\dot{\gamma}}(\theta,\varphi) d\theta$
of the strong shear regime
agree well with $P_{\rm J}(\varphi)$ with $l = 7.1$. 
We thus use $l=7.1$ for the present rigid rod, which is composed of
freely rotating beads.

\subsection{Analytic formula for the viscosity}

Hinch and Leal \cite{Hinch_Leal_factor, Leal_Hinch_K} studied the rheological
properties of a dilute dispersion of rigid non-spherical particles in
steady shear flow. 
They obtained an analytical formula for the dispersion viscosity. 
We analyze our numerical results with their formula. 
The dispersion viscosity $\eta$ is given by the ensemble average of the
temporal viscosity $\hat{\eta}(\theta, \varphi)$ using the PDF of the
two angles for the rotating rigid rods $P_{\dot{\gamma}}(\theta,
\varphi)$, which satisfies the normalization condition,
\begin{equation}
  \int_{-\frac{\pi}{2}}^{\frac{\pi}{2}} \cos \theta
   d \theta \int_{-\frac{\pi}{2}}^{\frac{\pi}{2}} d \varphi P_{\dot{\gamma}}(\theta, \varphi) = 1.
\end{equation}
The temporal shear viscosity is found to be
\begin{eqnarray}
  \hat{\eta}(\theta, \varphi) &=& \eta_f
   \left[1 +
\Phi \left( A \cos^4\theta \sin^2 2\varphi +
	 2B \cos^2 \theta + \frac{2}{I_3} \right.\right.\nonumber\\
  &&~~~~~~~~~~~~~~\left.\left. + \frac{D_{\rm r}}{\dot{\gamma}} F \frac{1}{2}\cos^2\theta
	      \sin 2\varphi \right)
   \right],
  \label{eta_formula}
\end{eqnarray}
where $\Phi$ is the volume fraction of suspended particles, $D_{\rm r}$
is the rotational diffusion constant, and $A,B,F,I_{3}$ are the shape
functions given in previous studies \cite{Hinch_Leal_factor,
batchelor_I, Leal_Hinch_K}. 
In the case of rigid rod, $A$, $B$, $F$, and $I_{3}$ are dependent only
on the aspect ratio $l$. 
At $l=7.1$, $A=8.44$, $B=0.06$, and $I_3=0.99$.

The shear viscosity of the dispersion is obtained by substituting
Eq.~(\ref{eta_formula}) into Eq.~(\ref{eta}). 
When we consider the strong shear case $D_{\rm r}\ll\dot{\gamma}$, we
can safely neglect the last term in Eq.~(\ref{eta_formula}). 
The dynamics of the angle $\varphi$ become decoupled from the angle
$\theta$ because the angle $\theta$ is sufficiently small for a large
$l$ \cite{Leal_Hinch_K}. 
Thus, we obtain the following formula,
\begin{eqnarray}
  \eta(\dot{\gamma})&=&~\int \hat{\eta}(\theta,\varphi)
   P''(\theta)P'(\varphi) d\theta d\varphi\\
  &=&~\eta_{\rm f} \left[ 1 +
 \Phi \biggl( A\langle\cos^4\theta\rangle_{\theta} \langle \sin^2 2\varphi\rangle_{\varphi} +
   2B\langle\cos^2 \theta\rangle_{\theta} \right.\biggr.\nonumber \\
&&~~~~~~~~~~~~~~\left.\left.+ \frac{2}{I_3} + \Delta E\right),
  \right],
   \label{eta_concrete_formula}
\end{eqnarray}
where $\langle f(\theta)\rangle_{\theta} =
\int_{-\frac{\pi}{2}}^{\frac{\pi}{2}} \cos \theta d \theta
f(\theta)P''(\theta)$, $\langle g(\varphi) \rangle_{\varphi} =
\int_{-\frac{\pi}{2}}^{\frac{\pi}{2}} d \varphi g(\varphi) P'(\varphi)$,
$P''(\theta)\equiv\int_{-\frac{\pi}{2}}^{\frac{\pi}{2}}
P_{\dot{\gamma}}(\theta,\varphi) d\varphi$, and $\Delta E$ is the error
arising from the separation of integrals over $\theta$ and $\varphi$. 
We can neglect $\Delta E$ safely because $\Delta E$ is sufficiently
small in comparison to the other terms.

\section{RESULTS}

\begin{figure}[htb]
\begin{center}
\includegraphics[width=1.0\hsize]{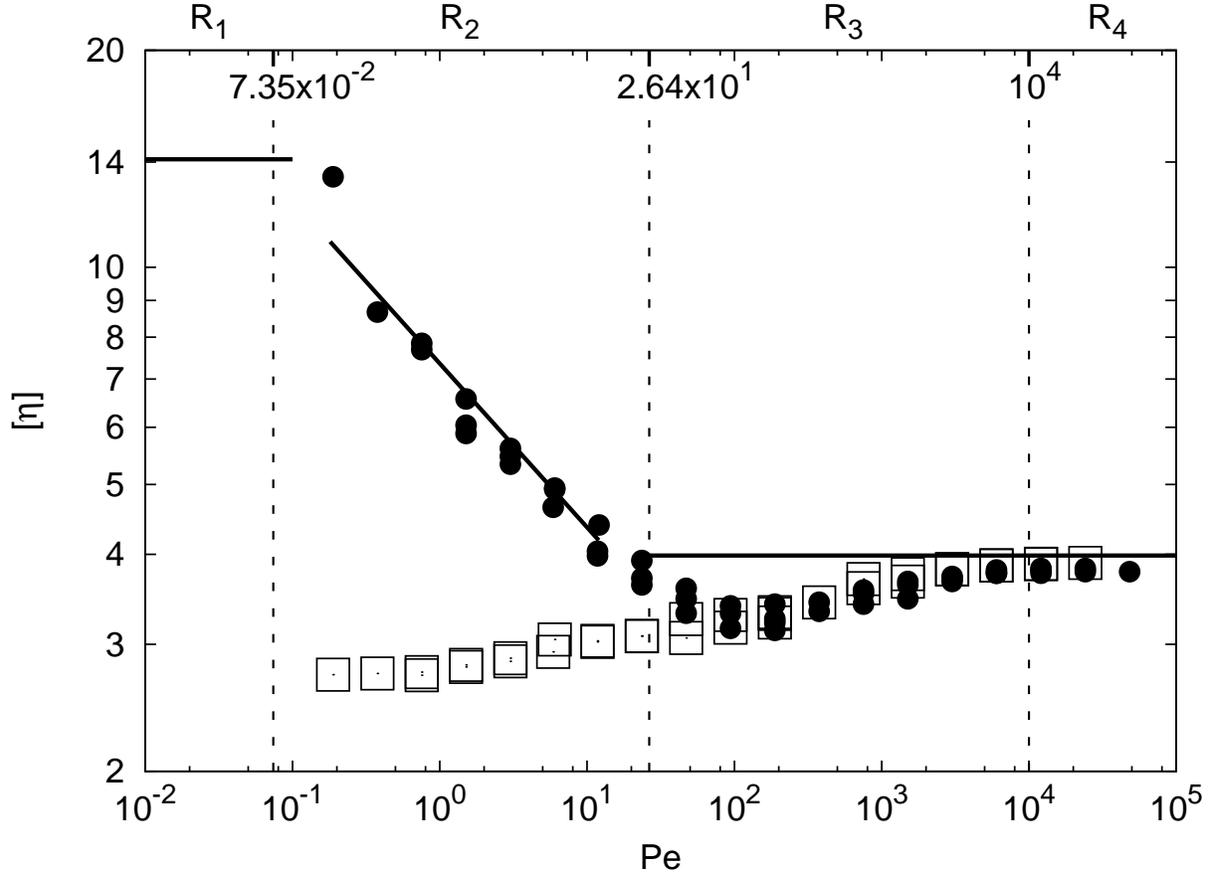}
\caption{The intrinsic viscosity as a function of {Pe}. $[\eta]$
 (circle) and $[\eta_{\theta}]$ (square). The three solid lines
 correspond to the theoretical result of Hinch and Leal
 \cite{Hinch_Leal_factor, Leal_Hinch_K}: the 1st Newtonian regime denoted by R$_1$, the shear-thinning regime denoted by R$_2$, and the 2nd Newtonian regime denoted by R$_3$ + R$_4$. In our simulation, the viscosity shows an undershoot before reaching the 2nd Newtonian regime R$_3$.}
\label{intrinsic_viscosity}
\end{center}
\end{figure}

In Fig.~\ref{intrinsic_viscosity}, we plotted the intrinsic viscosity
\begin{equation}
  [\eta]\equiv   \frac{\eta - \eta_{\rm f}}{\eta_{\rm f} \Phi}
    \label{def_intrinsic}
\end{equation}
of the dispersion obtained from the present simulations as a function of
the Peclet number {Pe}. 
{Pe} is the dimensionless number that represents the strength of the
shear flow normalized by that due to thermal fluctuations. 
In our work, {Pe} is defined as
\begin{equation}
  {\rm Pe} = \frac{6\pi\eta_{\rm f} \sigma^{3}\dot{\gamma}}{k_{\rm B}T}.
   \label{def_peclet}
\end{equation}
We find that the intrinsic viscosity $[\eta]$ gradually changes from
non-Newtonian (shear-thinning) to Newtonian behavior with increasing
Peclet number, as shown in Fig.~\ref{intrinsic_viscosity}. 
The present simulation data for $[\eta]$ show shear-thinning behavior
for ${\rm Pe} < 10^2$ and 2nd Newtonian behavior for $10^4 < {\rm Pe}$. 
Those results are in good agreement with previous theoretical studies
\cite{Hinch_Leal_factor, Leal_Hinch_K, Leal_Hinch_review}.

To quantitatively compare our results with those of Hinch and Leal
\cite{Hinch_Leal_factor, Leal_Hinch_K}, we obtain the relation between {Pe} in our
definition and $\dot{\gamma}/D_{\rm r}$, which is used in Hinch and
Leal's work \cite{Hinch_Leal_factor} instead of {Pe}. 
On the basis of the shell model \cite{dr1, dr2}, the rotational diffusion constant $D_{\rm r}$ for a rigid rod is calculated as 
\begin{equation}
  D_{\rm r}=\frac{3(\ln l + d (l))k_{\rm B}T}{\pi \eta_{\rm f} L^3},
   \label{def_dr}
\end{equation}
\begin{equation}
  d (l) = - 0.662 + \frac{0.917}{l} - \frac{0.05}{l^2}.
\end{equation}
In the shell model mentioned above, the contour of the macromolecules of
arbitrary shape is represented by a shell composed of many identical
small beads. 
The shell model can be adequately modeled by decreasing the size of the
beads. 
From Eq.~(\ref{def_peclet}) and Eq.~(\ref{def_dr}), the relation between $\dot{\gamma}/D_{\rm r}$ and {Pe} is expressed as
\begin{equation}
  \frac{\dot{\gamma}}{D_{\rm r}} = \frac{l^3}{18(\ln l + d (l))}{\rm Pe}.
\end{equation}
The theoretical model of Hinch and Leal is plotted also in
Fig.~\ref{intrinsic_viscosity} with the solid lines in the three
different regimes, namely, the weak (R$_{1}$), intermediate (R$_{2}$),
and strong (R$_{3}$ + R$_{4}$) shear regimes.

According to the work of Hinch and Leal \cite{Hinch_Leal_factor}, for
the weak shear regime $\dot{\gamma}/D_{\rm r} \ll 1$, namely, ${\rm Pe}
\ll 7.35\times10^{-2}$, which is denoted by R$_{1}$ in
Fig.~\ref{intrinsic_viscosity}, $[\eta]$ is constant. 
On the basis of Ortega's work \cite{de_la_rorre}, the intrinsic
viscosity $[\eta]$ of the weak-shear flow regime for a rigid rod with a
short aspect ratio is calculated as
\begin{equation}
    [\eta] = \frac{4}{15} \frac{l^2}{\ln l + \Upsilon(l)},
\end{equation}
\begin{equation}
  \Upsilon(l) = -0.90 - \frac{1.38}{l} + \frac{8.87}{l^2} - \frac{8.82}{l^3}.
\end{equation}
This expression is identical to Hinch and Leal's result in the limit of
$l \rightarrow \infty$. 
The lowest shear rate that we consider in the present simulations is
still not in the weak-shear regime because of the extremely long
simulation time needed to obtain reliable data.

For the intermediate shear regime $1\ll\dot{\gamma}/D_{\rm r} \ll l^3 +
l^{-3}$, namely, $7.35\times10^{-2} \ll {\rm Pe} \ll 26.4$, which is
denoted by R$_{2}$ in Fig.~\ref{intrinsic_viscosity}, the intrinsic
viscosity $[\eta]$ shows shear-thinning as derived from
Eq.~(\ref{eta_concrete_formula}),
\begin{equation}
  [\eta] = C_1 {\rm Pe}^{-1/3} + 2B + 2/I_3,
   \label{shear_thinning}
\end{equation}
where $C_1$ is an arbitrary constant. 
Figure~\ref{intrinsic_viscosity} shows good agreement between the data
from the present simulation data with that of Hinch and Leal, where
$C_1$ was determined to fit the simulation data. 
When $l$ is sufficiently large, the contributions from the last two
terms in Eq.~(\ref{shear_thinning}) become negligible, and $[\eta]
\propto {\rm Pe}^{-1/3}$.

For the strong shear regime $l^3 + l^{-3} \ll \dot{\gamma}/D_{\rm r}$,
namely, $26.4 \ll {\rm Pe}$, which is denoted by R$_{3}$ and R$_{4}$ in
Fig.~\ref{intrinsic_viscosity}, the theory predicts that the intrinsic
viscosity $[\eta]$ is constant at
$A \langle \sin^2 2\varphi \rangle_{J} + 2B + 2/I_3 = 3.99 = [\eta_\infty]$
from Eq. (\ref{eta_concrete_formula}).
Here $\langle \cdots \rangle_{J}$ denotes the ensemble average, which is
calculated as
\begin{equation}
  \langle f(\varphi) \rangle_{J} = \int_{-\frac{\pi}{2}}^{\frac{\pi}{2}} d \varphi f(\varphi) P_{\rm J}(\varphi).
\end{equation}
We obtained $[\eta_\infty]=3.82$ from our numerical data in the high
shear regime, which is  denoted by R$_{4}$ in
Fig.~\ref{intrinsic_viscosity}. 
The error from the theoretical value $3.97$ is within $4.26\%$.

For the regime $10^2<{\rm Pe}<10^3$, which is denoted by R$_{3}$ in
Fig.~\ref{intrinsic_viscosity}, the behavior of $[\eta]$ shows a notable
undershoot before reaching the high-shear limiting 2nd Newtonian
viscosity. 
This is attributable to the fluctuations in $\theta$. 
It gives rise to the deviations of $P''(\theta)$ from its high-shear
limiting form $P^{*}(\theta) \equiv  \int_{-\frac{\pi}{2}}^{\frac{\pi}{2}}
P_{J}(\theta,\varphi) d\varphi$ and increases with decreasing {Pe}. 
The rods tend to align in the flow direction with increasing {Pe}. 
Therefore, $\langle \cos^4 \theta \rangle$ and
$\langle \cos^2 \theta \rangle$
monotonically increase up to their high-shear limiting values with increasing shear rate. 
This leads to an increase in $[\eta]$ up to $[\eta_\infty]$ through
Eq.~(\ref{eta_concrete_formula}). 
To examine the role of thermal fluctuations in $\theta$ in more detail,
let us define
\begin{eqnarray}
   [\eta_{\theta}]  &=&
   A \langle\sin^2\varphi\rangle_{J}
   \langle\cos^4\theta\rangle_{\theta} +
   2B \langle\cos^2 \theta\rangle_{\theta} + \frac{2}{I_{3}}\\
   &=&
   1.81 \langle\cos^4\theta\rangle_{\theta} +
   0.12 \langle\cos^2 \theta\rangle_{\theta} + 2.02
   \label{eta_concrete_formula_NTN}
\end{eqnarray}
to estimate the contribution of $\theta$ fluctuations on the total
intrinsic viscosity of the dispersion. 
Here, $P(\varphi)=P_{\rm J}(\varphi)$ is assumed in
Eq.~(\ref{eta_concrete_formula}), and $\langle \cos^4 \theta
\rangle_{\theta}$ and $\langle \cos^2 \theta \rangle_{\theta}$ are
evaluated numerically from the present simulations. 
The results are plotted in Fig.~\ref{intrinsic_viscosity} with the
square symbols. 
One can see that the data of $[\eta_{\theta}]$ almost perfectly collapse
onto those of $[\eta]$ for $10^2<{\rm Pe}$.

\begin{figure}[htb]
\begin{center}
\includegraphics[width=1.0\hsize]{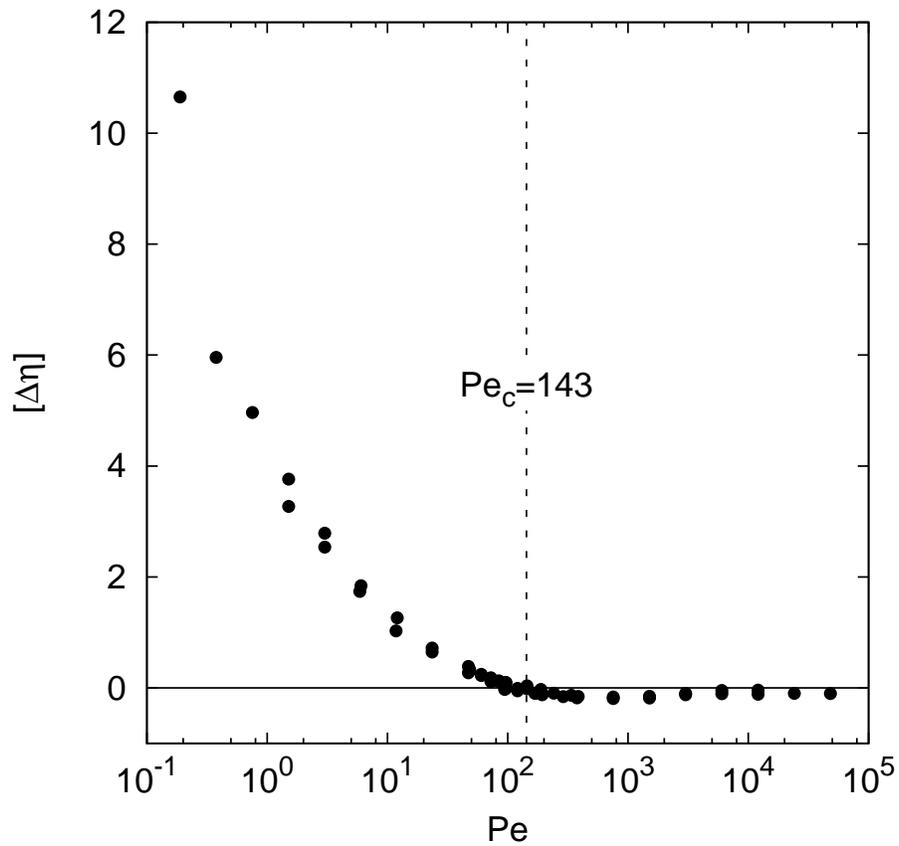}
\caption{The behavior of $[\Delta \eta]$ as a function of {Pe}. $[\Delta \eta]$ goes to zero around ${\rm Pe}\approx150$.}
\label{delta_intrinsic_viscosity}
\end{center}
\end{figure}

On the other hand, the shear-thinning behavior observed for $10^2>{\rm
Pe}$ is attributable to the effect of the thermal fluctuations in
$\varphi$. 
It gives rise to the deviations of $P'(\varphi)$ from its high-shear
limiting form $P_J(\varphi)$ and increases with decreasing {Pe}. 
To examine this effect quantitatively, we introduce 
\begin{eqnarray}
   [\Delta \eta] &\equiv& [\eta] - [\eta_{\theta}]\\
   &=& A\langle\cos^4\theta \rangle_{\theta}
   (\langle \sin^2 2\varphi \rangle_{\varphi} -
   \langle \sin^2 2\varphi \rangle_{J})
   \label{analytic_delta_eta}
\end{eqnarray} 
to eliminate the contribution of $\theta$ fluctuations from the total
intrinsic viscosity of the dispersion. 
Figure~\ref{delta_intrinsic_viscosity} shows the behavior of $[\Delta
\eta]$ as a function of {Pe}. 
$[\Delta \eta]$ decreases with increasing {Pe}, and finally $[\Delta
\eta]$ goes to zero around ${\rm Pe}\approx150$. 
This value is considerably different from the value ${\rm Pe}=26.4$
predicted by Hinch and Leal \cite{Hinch_Leal_factor} for the viscosity
transition from the shear-thinning to the 2nd Newtonian but agrees well
with our theoretical prediction of ${\rm Pe}_{\rm c}=143$, at which the
dynamical crossover from Brownian to non-Brownian behavior takes place
in the rotational motion of the rotating rod at $l=7.1$
\cite{dynamics_crossover}.

\section{DISCUSSION}

Let us discuss the numerical models with which the viscosity transition
to the 2nd Newtonian regime takes place based on
Eq.~(\ref{eta_concrete_formula}). 
For the strong shear regime $1 \ll \dot{\gamma}/D_{\rm r}$, we can
estimate $\langle\cos^2 \theta\rangle_{\theta}\simeq1$ and
$\langle\cos^4 \theta\rangle_{\theta}\simeq1$ because
$P_{\dot{\gamma}}(\theta, \varphi)\simeq\ P^{*}(\theta)P(\varphi)$. 
Here, $P(\varphi)$ satisfies the Fokker-Planck equation, shown as Eq.~(\ref{fokker-planck}), for which the formal solution is given by 
\begin{eqnarray} P(\varphi) = C_1\int^{\pi}_0 d\psi
 \exp \left(
	 - \frac{\dot{\gamma}}{4D_r} f(\psi, \varphi)
	\right), \label{solution_focker_jeffrey} \\
  f(\psi, \varphi) = \psi - (1-\frac{2}{l^2+1})\sin \psi \cos (\psi-2\varphi),
\end{eqnarray}
where $C_1$ is determined from the normalization condition,
$\int_{-\frac{\pi}{2}}^{\frac{\pi}{2}} P(\varphi) d\varphi = 1$. 
When $\dot{\gamma}/D_{\rm r}$ is sufficiently large, $P(\varphi)$
converges to $P_{\rm J}(\varphi)$ represented by Eq.~(\ref{p_j}), and
the viscosity displays 2nd Newtonian behavior.

\begin{figure}[htb]
\begin{center}
\includegraphics[width=1.0\hsize]{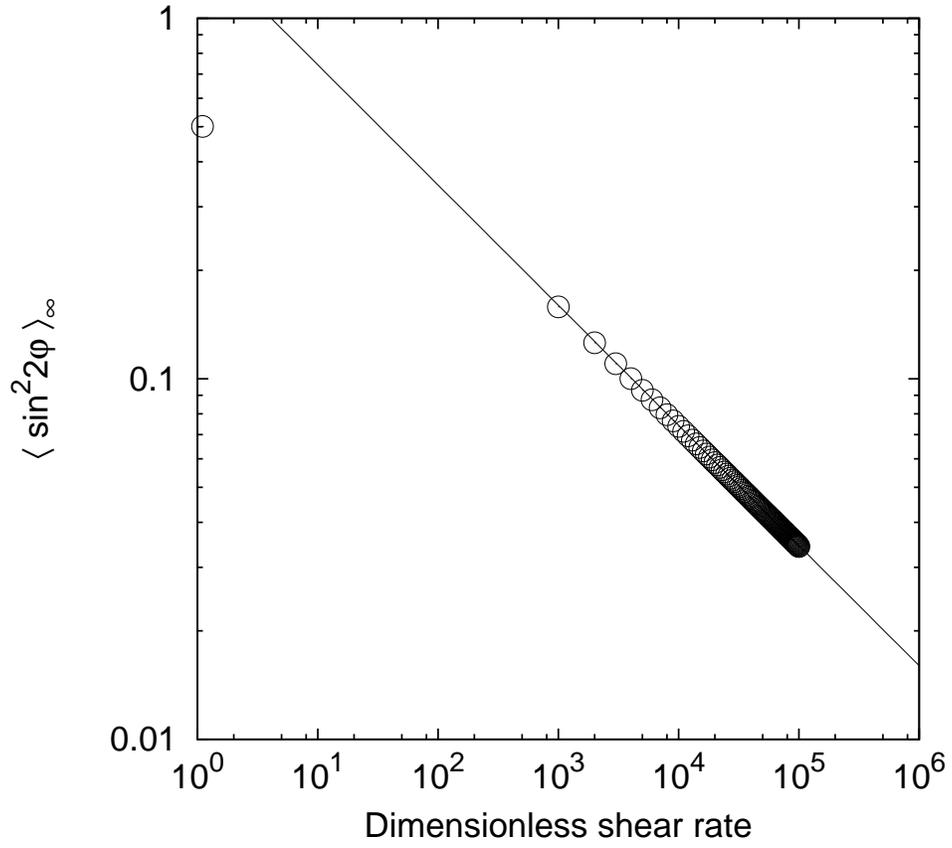}
\caption{The behavior of $\langle \sin^{2} 2\varphi \rangle_{\infty}$ as a function of $\dot{\gamma}/D_r$. Numerical results (circle). The solid line corresponds to $\left( \dot{\gamma}/D_{r}  \right)^{-1/3}$} \label{eta_with_using_Pinf}
\end{center}
\end{figure}

\begin{figure}[htb]
\begin{center}
\includegraphics[width=1.0\hsize]{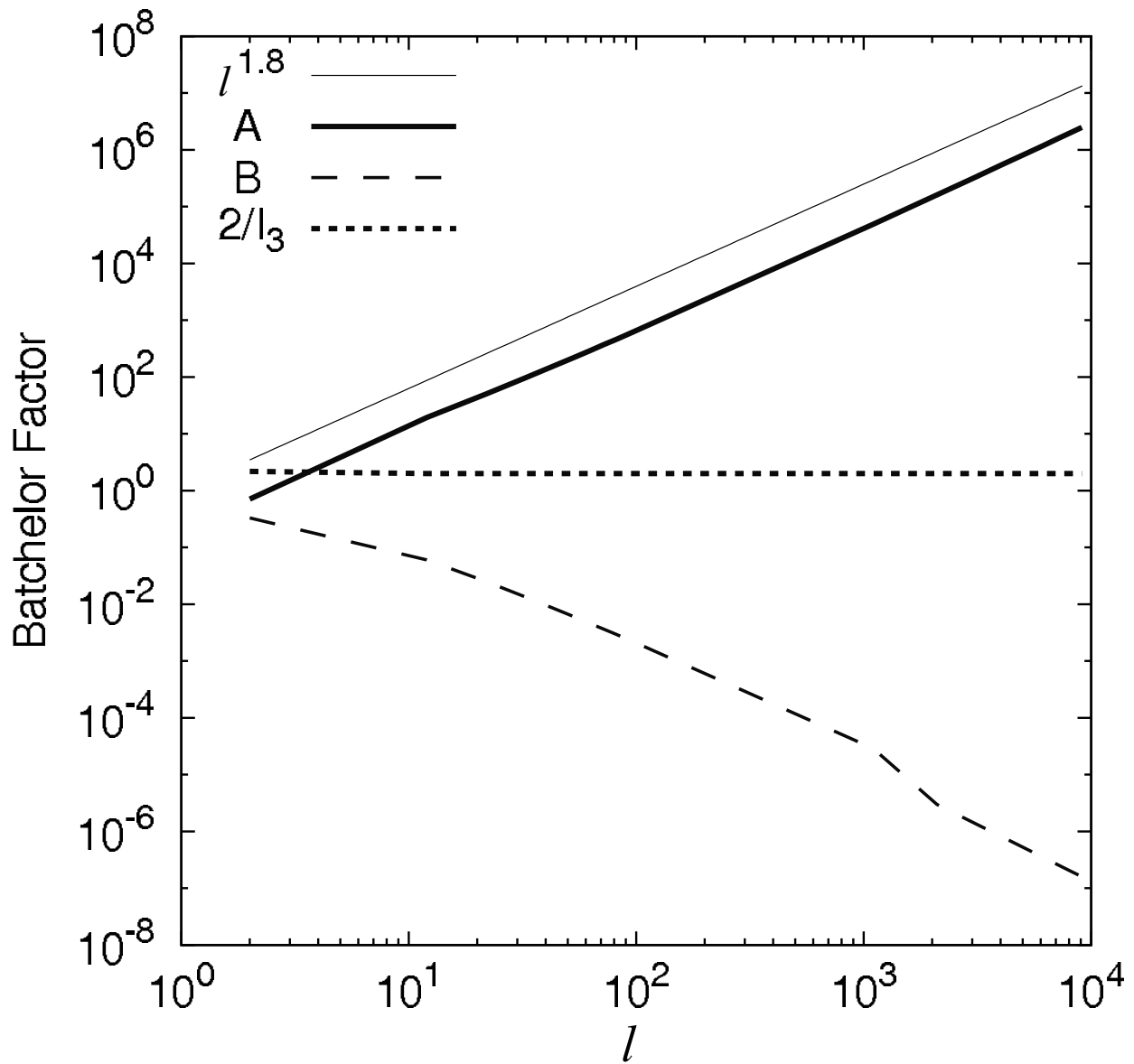}
\caption{The behavior of A, B, and $2/I_{3}$ as a function of the aspect ratio $l$. A (bold solid line), B (dashed line), $2/I_{3}$ (dotted line). The thin solid line represents $l^{1.8}$.}
\label{batchelor_factor}
\end{center}
\end{figure}

The above discussion is not valid in the limit of $l \rightarrow
\infty$, which corresponds to an infinitely-long or equivalently
infinity-thin rod. 
In this limit, the angular velocity of the tumbling rod becomes zero at
$\varphi=0$ from Eq.~(\ref{omega_velocity}). 
Thus, the rod cannot continue rotational motion without thermal
fluctuations. 
This is because the hydrodynamic torque acting on the rod becomes zero
at $\varphi=0$ for $l \rightarrow \infty$. 
Therefore, $P(\varphi)$ in Eq.~(\ref{solution_focker_jeffrey}) is modified to
\begin{eqnarray}
  P_{\infty}(\varphi) = C_2\int^{\pi}_0 d\psi
   \exp \left(
	 - \frac{\dot{\gamma}}{4D_r} \left[ \psi - \sin \psi \cos
				      (\psi-2\varphi) \right]
	\right), \label{solution_infinity}
\end{eqnarray}
where $C_2$ is determined from the normalization condition,
$\int_{-\frac{\pi}{2}}^{\frac{\pi}{2}} P_{\infty}(\varphi) d\varphi =1$. 
Using Eq.~(\ref{solution_infinity}), the intrinsic viscosity $[\eta]$ is rewritten as
\begin{equation}
   [\eta] = A \langle \sin^2 2\varphi \rangle_{\infty} + 2B + 2/I_{3},
   \label{eta_inf}
\end{equation}
\begin{equation}
  \langle f(\varphi)\rangle_{\infty} = \int_{-\frac{\pi}{2}}^{\frac{\pi}{2}} d \varphi
   f(\varphi)P_{\infty}(\varphi).
\end{equation}
It is demonstrated in Fig.~\ref{eta_with_using_Pinf} that the first term
in Eq.~(\ref{eta_inf}) shows $\langle \sin^2 2\varphi
\rangle_{\infty}\propto(\dot{\gamma}/D_{\rm r})^{-1/3}$ for the entire
range of Pe. 
Figure \ref{batchelor_factor} shows that only $A$ is increasing with
increasing $l$, while $B$ and $2/I_3$ tend to be decreasing or constant
upon increasing $l$. 
We estimate $A/(2B + 2/I_{3})\sim l^{1.8}$ for $l\rightarrow\infty$. 
This indicates that 
\begin{equation}
[\eta]\propto\langle \sin^2 2\varphi
 \rangle_{\infty}\propto(\dot{\gamma}/D_{\rm r})^{-1/3}
\end{equation}
holds for the entire range of Pe without indicating the occurrence of
2nd Newtonian behavior. 
The same conclusion can be derived by considering a characteristic shear
rate $\dot{\gamma}^*$ at which the first term in Eq.~(\ref{eta_inf})
becomes comparable to the remaining terms. 
The condition is satisfied at 
\begin{equation}
\frac{\dot{\gamma}^*}{D_{\rm r}}  \sim l^{5.4}.
   \label{condition_Pe}
\end{equation}
This indicates $\dot{\gamma}^*\rightarrow\infty$ for $l\rightarrow\infty$.

In the case of a previous numerical study \cite{shear_thinning_sim2},
the hydrodynamic force acting on each bead particle was considered via
the RPY tensor. 
Although the translational hydrodynamic force was properly considered,
the rotational hydrodynamic torque acting on each bead particle was
completely ignored in that study. 
Therefore, it is suspected that the hydrodynamic torque acting on the
rod becomes zero at $\varphi=0$. 
Therefore, the rod cannot continue rotational motion at a high shear
rate, where the effect of thermal fluctuations disappears. 
This situation is exactly the same as the case of $l\rightarrow\infty$. 
We expect that the 2nd Newtonian regime could be correctly reproduced
with the RPY tensor approach if the hydrodynamic torque is taken into
account properly.

\section{CONCLUSION}

In the present study, we numerically calculated the intrinsic viscosity
$[\eta]$ of a dilute dispersion of rigid rods using a DNS method known
as SPM. 
Simulations were conducted under the influence of thermal fluctuations
and shear flow in the ranges of $5.0\times10^{-4}<k_{\rm B}T<32$ and
$5.0\times10^{-3}<\dot{\gamma}<2.0\times10^{-2}$, respectively. 
We have successfully reproduced the viscosity transition from the
shear-thinning to the 2nd Newtonian regimes, as was correctly predicted
by the theoretical model of Hinch and Leal \cite{Hinch_Leal_factor, Leal_Hinch_K}.

There are, however, some discrepancies between the theoretical
predictions and the results of the present simulations. 
By defining $[\Delta \eta]$ to eliminate the effects of fluctuations in
$\theta$, which is not considered in the theoretical model, we confirmed
that the viscosity transition from the shear-thinning to the 2nd
Newtonian takes place around ${\rm Pe}=150$. 
This value is considerably larger than the value of $26.4$ predicted by
Hinch and Leal \cite{Hinch_Leal_factor} but agrees well with our
theoretical prediction of ${\rm Pe}_{\rm c}=143$, at which the dynamical
crossover from Brownian to non-Brownian behavior takes place in the
rotational motion of the rotating rod \cite{dynamics_crossover}.

We have analyzed the mechanism of the viscosity undershoot observed in
our simulation before reaching the 2nd Newtonian regime. 
Shear flow suppresses fluctuations in $\varphi$ and $\theta$ as its rate
is increased. 
The former contributes to decrease [$\eta$], but the later contributes
to increase [$\eta$]. 
The undershoot takes place because of the two competing effects.

We also conclude that the viscosity transition to the 2nd Newtonian
regime can be reproduced correctly only if the hydrodynamic torque is
properly taken into account in numerical models of the dispersions.

\section*{ACKNOWLEDGMENTS}

The authors would like to express their gratitude to Dr.~T.~Murashima and
Dr.~Y.~Nakayama for useful comments and discussions.

\newpage

\end{document}